\documentclass[a4paper, 10 pt, conference]{ieeeconf}
\usepackage{amsfonts,amsmath,amssymb} % assumes amsmath package installed

\DeclareMathOperator{\diag}{diag}         % Define diag operator
\def\diag{{\rm diag}}
\def\begce{\begin{center}}
\def\endce{\end{center}}
\def\begar{\begin{array}}
\def\endar{\end{array}}
\def\begeq{\begin{equation}}
\def\endeq{\end{equation}}
\def\begdi{\begin{displaymath}}
\def\enddi{\end{displaymath}}
\def\begdis{\begin{eqnarray*}}
\def\enddis{\end{eqnarray*}}
\def\begeqa{\begin{eqnarray}}
\def\endeqa{\end{eqnarray}} 
\usepackage{psfrag,color}
\usepackage{enumerate,cite,latexsym,graphicx}
\newtheorem{theorem}{Theorem}

\newtheorem{definition}{Definition}

\title{\LARGE \bf  A Survey of Riccati Equation Results in Negative Imaginary Systems Theory and Quantum Control Theory}

\author{Ian R.~Petersen        
  \thanks{This work was supported by the
Australian Research Council (ARC) under grant FL110100020 and the Air Force Office of Scientific
Research (AFOSR), under agreement number FA2386-16-1-4065. }%
\thanks{Ian R. Petersen is with the Research School of  Engineering, The Australian National University, Canberra, ACT 2601, Australia.
         {\tt\small i.r.petersen@gmail.com} } 
}%              
\begin{document}

\maketitle
\thispagestyle{empty}
\pagestyle{empty}

\begin{abstract}
This paper presents a survey of some new applications of algebraic Riccati equations. In particular, the paper surveys some recent results on the use of algebraic Riccati equations in testing whether a system is negative imaginary and in synthesizing state feedback controllers which make the closed loop system negative imaginary. The paper also surveys the use of Riccati equation methods in the control of quantum linear systems including coherent $H^\infty$ control.  
\end{abstract}

%%%%%%%%%%%%%%%%%%%%%%%%%%%%%%%%%%%%%%%%%%%%%%%%%%%%%%%%%%%%%%%%%%%%%%%%%%%%%%%%
\section{Introduction} \label{sec:intro}
The algebraic Riccati equation has been applied in a range of applications in modern control theory including the linear quadratic regulator problem, the linear quadratic Gaussian problem and the $H^\infty$ control problem; e.g., see \cite{KS72,ZDG96,PAJ91}. In particular, it has played a major role in robust control theory; e.g., see \cite{ZDG96,PUSB}. In this paper, we survey some results in which the Riccati equation plays a role in the emerging areas of negative imaginary systems theory and quantum linear systems theory. 

The theory of negative imaginary systems is an emerging theory which is attracting  interest among control theory researchers; e.g., see \cite{LP06,XiPL1,LP_CSM,XiPL2,XiLP1,OPM11,BV13,FN13,LX15,LX16,FLN16,BFF16,DPS16}. This theory is broadly applicable to problems of robust vibration control for flexible structures; e.g., see \cite{LP06,LP_CSM,CH10a,MKPL11}. Such flexible structures can be modelled by high order linear systems models with highly resonant dynamics \cite{HM01,PMS02,PRE02}. A particular problem in the control of such systems is the fact that unmodelled spillover dynamics can severely degrade control system performance or lead to instability if the controller is not designed to be robust against this type of uncertainty; e.g., see  \cite{BM90,DLSS08}. In addition, uncertainties in resonant frequencies and damping levels can cause similar problems of poor control system performance or instability. Negative imaginary systems theory provides a way of analyzing robustness and designing robust controllers for such flexible structures in the case of collocated force actuators and position sensors; e.g., see \cite{BMP1,MM05,MVB06,CH10a,MMB11,MKPL1,MLKP1,MKPL5,MKPL11,XiLaP1}. Although much of the theory of negative imaginary systems revolves around the use of linear matrix inequalities (LMIs), a number of recent results have emerged in which Riccati equations are used instead of LMIs; see \cite{MKPL11}. 

Quantum
feedback control systems have been an active area of research in recent years; e.g., see
\cite{BEL83,WIS94,DJ99,YK03a,YK03b,DP3}. In particular, 
there has been considerable interest in the feedback
control and modeling of
linear quantum systems; e.g., see
\cite{DJ99,YK03a,YK03a,JNP1,NJP1,GGY08,MaP3,MaP4,MAB08,YNJP1,PET08A,ShP5,NJD09,GJ09,GJN10,WM10}.
 Linear quantum  
systems commonly arise in the area of quantum optics; e.g., see
\cite{WM94,GZ00,BR04}.  Feedback control of quantum systems aims to achieve
closed loop properties such as stability \cite{PUJ2},
robustness \cite{JNP1} and entanglement \cite{YNJP1,NPJ1}. 

 An important class of linear quantum stochastic models describe the Heisenberg evolution of the (canonical) position and momentum, or annihilation and creation operators of several independent open quantum harmonic oscillators that are coupled to external coherent bosonic fields; e.g.,  see \cite{WM94}, \cite{WM10}, \cite{GZ00}, \cite{BE08}, \cite{WD05,JNP1,NJP1,NJD09,YAM06,MAB08,YNJP1,GGY08,SSM08,GJN10}).
These linear stochastic models describe quantum optical devices such as optical cavities \cite{BR04}, \cite{WM94}, linear quantum amplifiers \cite{GZ00}, and finite bandwidth squeezers \cite{GZ00}.  In particular, we consider  linear quantum stochastic differential equations
driven by quantum Wiener processes; see \cite{GZ00}. Further details on
quantum stochastic differential equations and quantum  Wiener
processes can be found in  \cite{HP84,PAR92,BHJ07}.  

   Some recent papers on the feedback control of linear quantum systems 
have considered the case in which the feedback controller  itself is
also a
quantum system. Such feedback control is often
referred to as coherent quantum control; e.g., see
\cite{WM94a,SL00,YK03a,YK03b,JNP1,NJP1,MaP1a,MaP3,MaP4,MAB08,GW09}. In particular, some recent results on the coherent $H^\infty$ control problem and the coherent LQG control problem involve the use of the Riccati equation and in this paper, we survey some of these results; e.g., see \cite{JNP1,PET10Ba,PET13Da}.

\section{The Riccati Equation in Negative Imaginary Systems Theory}
Negative imaginary (NI) systems theory is an emerging area of robust control theory which is concerned with the analysis and robust control of flexible systems with co-located force actuators and position sensors. 

\begin{definition}\cite{LP06,XiPL1,MKPL10}\label{Def:NI}
A square transfer function matrix $M(s)$ is NI if  the following
conditions are satisfied:
\begin{enumerate}
\item $M(s)$ has no pole in $Re[s]>0$.
\item For all $\omega >0$ such that $s=j\omega$ is not a pole of $M(s
)$,
\begin{equation}\label{eq:NI:def}
    j\left( M(j\omega )-M(j\omega )^{\ast }\right) \geq 0.
\end{equation}
\item If $s=j\omega _{0}$ with $\omega _{0}>0$ is a pole of $M(s)$, then it is a simple pole and the residue matrix $K=\underset{%
s\longrightarrow j\omega _{0}}{\lim }(s-j\omega _{0})jM(s)$ is
Hermitian and  positive semidefinite.
 \item If $s=0$ is a pole of $M(s)$, then
$\underset{s\longrightarrow 0}{\lim }s^{k}M(s)=0$ for all $k\geq3$
and $\underset{s\longrightarrow 0}{\lim }s^{2}M(s)$ is Hermitian and
positive semidefinite.
\end{enumerate}
\end{definition}

The following definition defines the strict negative imaginary property which is needed in the NI stability result. 

\begin{definition}[See \cite{LP06}.]
  \label{dfn:2}
  A square real-rational proper transfer function matrix $N(s)$ is
  termed \emph{strictly negative imaginary} (SNI) if
  \begin{enumerate}
  \item $N(s)$ has no poles in $\Re[s]\ge0$;
  \item $j[N(j\omega)-M^*(j\omega)]>0$ for $\omega\in(0,\infty)$.
  \end{enumerate}
\end{definition}

We now present the main  stability result  of negative imaginary systems theory that guarantees the
robustness and stability  of control systems involving the positive-feedback interconnection of an NI system
and an SNI system; see also \cite{LP06,XiPL1,LP_CSM}.

\begin{theorem}[\cite{LP06,XiPL1}]
  \label{thm:2}
Consider an NI transfer function matrix
  $M(s)$ with no poles at the origin and an SNI transfer
  function matrix $N(s)$, and suppose that $M(\infty)N(\infty)=0$
  and $N(\infty)\geq 0$. Then, the positive-feedback interconnection
  of $M(s)$ and $N(s)$
  is internally stable if and only if
\begin{equation}
\label{DC_gain}
  \lambda_{max}(M(0)N(0))<1.
\end{equation}
\end{theorem}
\ \\

An important result in the theory of negative imaginary systems is the following result which is referred to as the negative-imaginary lemma. This lemma can be used for testing if a given system is negative imaginary. It is also used in the proof of the above stability theorem.

\begin{theorem}[Negative Imaginary Lemma,\cite{LP06,XiPL1,MKPL11}.]\label{NI-lemma}
 Let $(A,B,C,D)$ be a minimal state-space realization of an $m \times m$
  real-rational proper transfer function matrix $M(s)$, where
  $A\in\mathbb{R}^{n \times n}$, $B\in\mathbb{R}^{n \times m}$,
  $C\in\mathbb{R}^{m \times n}$, $D\in\mathbb{R}^{m \times m}$. 
 Then, $M(s)$ is NI if and only if $D=D^T$ and there exist
matrices $P=P^{T}\geq 0$,
 $W\in \mathbb{R}^{m \times m}$, and $L\in \mathbb{R}^{m \times n}$
 such that the following linear matrix inequality (LMI) is
satisfied:
\begin{small}
\begin{align}\label{LMI:PR}
\begin{bmatrix}
PA+A^{T}P & PB-A^{T}C^{T} \\
B^{T}P-CA & -(CB+B^{T}C^{T})%
\end{bmatrix}%
 =
\begin{bmatrix}
-L^{T}L & -L^{T}W \\
-W^{T}L & -W^{T}W%
\end{bmatrix}%
\leq 0.
\end{align}
\end{small}
\end{theorem}

The following result is our first result in the theory of negative imaginary systems which uses the Riccati equation. This result is an alternative version of the negative imaginary lemma which uses the Riccati equation rather than the LMI (\ref{LMI:PR}). 

\begin{theorem}[\cite{MKPL11}]\label{reccati:ni}
Suppose the transfer function matrix  $%
M(s)$ has a minimal realization $
\begin{bmatrix}
\begin{array}{c|c}
A & B \\ \hline C & D
\end{array}
\end{bmatrix}$ such that  $CB+B^{T}C^{T}>0$. Then $M(s)$ is NI
if and only if $D=D^T$ and there exists a matrix $P\geq 0$ that
solves the  algebraic
Riccati equation%
\begin{equation}
PA_{0}+A_{0}^{T}P+PBR^{-1}B^{T}P+Q=0  \label{main rectta}
\end{equation}
 where%
\begin{align*}
A_{0} &=A-BR^{-1}CA, \\
R &=CB+B^{T}C^{T}, \text{ and  } \\
Q &=A^{T}C^{T}R^{-1}CA.
\end{align*}
\end{theorem}
\ \\

In our next result, we use an algebraic Riccati equation to synthesize a state feedback controller such that the closed loop system in NI. Consider the following  state space representation  for a  linear uncertain system  given as follows;
\begin{align}
\dot {x} &=Ax+B_{1}w+B_{2}u, \notag \\
z &=C_{1}x, \label{syst}\\
 W(s) &=\Delta(s) Z(s),  \notag
\end{align}%
where we  assume that $ W(s)$ and $Z(s)$ are  the Laplace transforms of the  signals $w(t)$ and $z(t)$.

Here, $A \in \mathbb{R}^{n \times n},B_1 \in \mathbb{R}^{n \times
m}, B_2 \in \mathbb{R}^{n \times r} ,C_1 \in \mathbb{R}^{m \times
n},$ and $\Delta(s) $ is an strictly negative imaginary uncertainty transfer function
matrix; e.g., see \cite{LP06,MKPL10}.  Also, suppose that $K\in \mathbb{R}^{r \times n}$ is a  state feedback matrix such that
$u=K x$. Then the closed-loop interconnection of  the system
\eqref{syst} with the state feedback control law is
given by

\begin{align}
\overset{\cdot }{x} &=(A+B_{2}K)x+B_{1}w,  \notag \\
z &=C_{1}x. \label{clsed_loop_syst}
\end{align}
and
\begin{align}
 W(s) =\Delta(s) Z(s).\label{syst200}
\end{align}%
Our aim  is to construct  the  matrix  $K$ such that the
corresponding closed-loop system \eqref{clsed_loop_syst} is stable
and satisfies the NI property.  From this, it follows  that  the
closed-loop uncertain system will be stable for any SNI uncertainty
$\Delta(s)$ \cite{LP06,LP_CSM} providing the corresponding  DC gain
condition (\ref{DC_gain}) is satisfied.

Consider the following Schur transformation:
\begin{align}
A_f&=U^T(A-B_{2}(C_{1}B_{2})^{-1}C_{1}A)U=\begin{bmatrix}
A_{11} & A_{12} \\
0 & A_{22}%
\end{bmatrix},  \label{AF_matrix} \\
B_f&=U^T(B_{2}(C_{1}B_{2})^{-1}-B_{1}R^{-1})=\begin{bmatrix}
B_{f1} \\
B_{f2}%
\end{bmatrix}, \label{Bf_matrix}  \\
\tilde{B}_1&=U^TB_{1}=%
\begin{bmatrix}
B_{11} \\
B_{22}%
\end{bmatrix}, \label{B1_matrix}
\end{align}
 where $U$ is a unitary matrix.

The transformation  \eqref{AF_matrix} can be constructed   such that
$A_{11}$ has all of its eigenvalues in the closed left half of the
complex plane and $A_{22}$ has all of its  eigenvalues in the open right half of the complex plane; i.e., $A_{22}$ is an  anti-stable matrix.

\begin{theorem}[\cite{MKPL11}]\label{syn:cont}
Given an uncertain system \eqref{syst} with
$C_{1}B_{2}$ non-singular  and $R=C_{1}B_{1}+B_{1}^{T}C_{1}^{T}>0$, define $A_f$, $B_f$, $\tilde{B}_1$, $U$, $A_{11}$, $A_{12}$, $A_{22}$, $B_{f1}$, $B_{f2}$, $B_{11}$ and $B_{22}$ as in \eqref{AF_matrix}-\eqref{B1_matrix}, where  $A_{22}$ is the anti-stable block of the $A_f$ matrix.
Then there exists a static state-feedback matrix  $K$ such that the closed-loop
system  (\ref{clsed_loop_syst}) is NI if  there exist  matrices
$T\geq 0$ and $S\geq 0$ such that \
\begin{align}
-A_{22}T-TA_{22}^{T}+B_{f2}RB_{f2}^{T} &=0, \label{lep111} \\
-A_{22}S-SA_{22}^{T}+B_{22}R^{-1}B_{22}^{T} &=0. \label{lep222}
\end{align}%
and $T-S>0$. Furthermore, a static state-feedback matrix $K$ which makes the closed-loop system (\ref{clsed_loop_syst}) NI and stabilizes the anti-stable matrix $A_{22}$ is
given by
\begin{equation} \label{k_matrix}
K=(C_{1}B_{2})^{-1}(B_{1}^{T}P-C_{1}A-R(B_{2}^{T}C_{1}^{T})^{-1}B_{2}^{T}P),
\end{equation} where $P=UP_{f}U^T$ and $P_{f}=\begin{bmatrix}
0 & 0 \\
0 & (T-S)^{-1}%
\end{bmatrix}\geq0$ satisfies the  algebraic Riccati equation
\begin{equation}
P_{f}A_{f}+A_{f}^{T}P_{f}-P_{f}B_{f}RB_{f}^{T}P_{f}+P_{f}\tilde{B}_{1}R^{-1}\tilde{B}_{1}^{T}P_{f}=0.
\label{reccta01}
\end{equation}%
\end{theorem}

\section{The Riccati Equation in Linear Quantum Systems Theory}
We consider a class of linear quantum system
models. These linear quantum system models take
the form of quantum stochastic differential equations (QSDEs) which are
derived from the quantum harmonic oscillator; e.g., see \cite{JNP1,PET10Ba,PET13Da}. We will survey a number of results involving Riccati equations which arise in the theory of such quantum linear systems. 
\subsection{Quantum Harmonic Oscillators}
\label{subsec:harmonic_oscillator}
We begin by considering a collection of $n$ independent quantum
harmonic oscillators which are
defined on a Hilbert space $\mathcal{H} =
L^2(\mathbb{R}^n,\mathbb{C})$; e.g., see
\cite{MEY95,PAR92,GJN10}. Elements of the Hilbert space $\mathcal{H}$,
$\psi(x)$ are the standard complex valued wave functions arising in
quantum mechanics where $x$ is a spatial variable.  Corresponding to this collection
of harmonic oscillators is a vector of  {\em annihilation operators}
$
%\label{anninil} 
a = \left[\begin{array}{cccc}a_1&a_2& \ldots & a_n
\end{array}\right]^T.
$
The adjoint of the operator $a_i$ is denoted $a_i^*$
and is referred to as a creation operator.

The quantum harmonic oscillators described above are assumed to be
coupled to $m$ external independent quantum fields modelled by bosonic
annihilation field operators $\mathcal{A}_1(t),
\mathcal{A}_2(t),\ldots,\mathcal{A}_m(t)$ which are defined on
separate Fock spaces $\mathcal{F}_i$ defined  over
$L^2(\mathbb{R})$ for each field operator
\cite{HP84,PAR92,BHJ07,NPJ1}. 
For each annihilation field operator
$\mathcal{A}_j(t)$, there is a corresponding creation field operator
$\mathcal{A}_j^*(t)$, which is defined on the same Fock space and is
the operator adjoint of $\mathcal{A}_j(t)$. 
The field annihilation operators are also collected
 into a vector of
operators defined as follows:
$
 \mathcal{A}=\left[\begin{array}{cccc}
\mathcal{A}_1& \mathcal{A}_2& \ldots & \mathcal{A}_m
\end{array}\right]^T.
$

In order to describe the joint evolution of the quantum harmonic
oscillators and quantum fields, we first specify the {\em Hamiltonian
operator} for the quantum system which is a Hermitian operator on
$\mathcal{H}$ of the form
\begin{equation}
\label{Hamiltonian}
{\bf H} = \frac{1}{2}\left[\begin{array}{cc}a^\dagger &
      a^T\end{array}\right]M
\left[\begin{array}{c}a \\ a^\#\end{array}\right]
\end{equation}
where $M \in \mathbb{C}^{2n\times 2n}$ is a Hermitian matrix of the
form
\begin{equation}
\label{tildeMN}
M= \left[\begin{array}{cc}M_1 & M_2\\
M_2^\# &     M_1^\#\end{array}\right]
\end{equation}
and $M_1 = M_1^\dagger$, $M_2 = M_2^T$. Also, we specify the {\em coupling operator} for
the quantum system to be an operator of the form
\begin{equation}
\label{coupling_operator}
L = \left[\begin{array}{cc}N_1 & N_2 \end{array}\right]
\left[\begin{array}{c}a \\ a^\#\end{array}\right]
\end{equation}
where $N_1 \in \mathbb{C}^{m\times n}$ and $N_2 \in
\mathbb{C}^{m\times n}$. Also, we write
\[
\left[\begin{array}{c}L \\ L^\#\end{array}\right] = N
\left[\begin{array}{c}a \\ a^\#\end{array}\right] =
\left[\begin{array}{cc}N_1 & N_2\\
N_2^\# &     N_1^\#\end{array}\right]
\left[\begin{array}{c}a \\ a^\#\end{array}\right].
\]
In addition, we define a {\em scattering
  matrix} which is a unitary matrix $S \in \mathbb{C}^{n\times
  n}$. 

The Heisenberg evolution of the operator vectors $a$, and $a^\dagger$ is described the following  QSDEs (e.g., see equations (1) and (2) in \cite{EMPUJ3a} and equations (7) and (9) in \cite{ShP5}): 
\begin{eqnarray}
\label{QSDE1}
\lefteqn{d \left[\begin{array}{l}
      a\\a^\#\end{array}\right]=} \nonumber \\
&& 
-\imath\left[ 
\left[\begin{array}{l}a\\a^\#\end{array}\right]
,{\bf H}\right]dt 
+\frac{1}{2}\left(L^\dagger\left[ 
\left[\begin{array}{l} a\\a^\#\end{array}\right]
,L^T\right]^T\right)^Tdt \nonumber \\
&& +\frac{1}{2}\left[L^\#,
\left[\begin{array}{l} a\\a^\#\end{array}\right]^T \right]^TLdt 
%\nonumber \\&& 
+ \left[
\left[\begin{array}{l}a\\a^\#\end{array}\right]
,L^T\right]d\mathcal{A}^\#\nonumber \\
&&- \left[
\left[\begin{array}{l}a\\a^\#\end{array}\right]
,L^\dagger\right]d\mathcal{A}.
\end{eqnarray}

Using (\ref{Hamiltonian}) and (\ref{coupling_operator}), these equations then lead to the following QSDEs, which describe the dynamics of the quantum system under consideration:
\begin{eqnarray}
\label{qsde3}
\left[\begin{array}{l} da(t)\\da(t)^\#\end{array}\right]
 &=& 
F\left[\begin{array}{l} a(t)\\a(t)^\#\end{array}\right]dt 
+ G \left[\begin{array}{l} d\mathcal{A}(t)
\\ d\mathcal{A}(t)^{\#} \end{array}\right];  \nonumber \\
\left[\begin{array}{l} d\mathcal{A}^{out}(t)
\\ d\mathcal{A}^{out}(t)^{\#} \end{array}\right] &=& 
H \left[\begin{array}{l} a(t)\\a(t)^\#\end{array}\right]dt 
+ K \left[\begin{array}{l} d\mathcal{A}(t)
\\ d\mathcal{A}(t)^{\#} \end{array}\right],\nonumber \\
\end{eqnarray}
where
\begin{eqnarray}
\label{FGHKform}
F &=& \left[\begin{array}{cc}F_1  & F_2\\
F_2^\# & F_1^\#\end{array}\right]; ~~
G = \left[\begin{array}{cc}G_1  & G_2\\
G_2^\# & G_1^\#\end{array}\right]; \nonumber \\
H &=& \left[\begin{array}{cc}H_1  & H_2\\
H_2^\# & H_1^\#\end{array}\right]; ~~
K = \left[\begin{array}{cc}K_1  & K_2\\
K_2^\# & K_1^\#\end{array}\right]. ~~
\end{eqnarray}
Here,
\begin{eqnarray}
\label{FGHKmatrices}
 F &=& -i J M -\frac{1}{2}  J N^\dagger J N; \nonumber \\
 G &=& -  J N^\dagger 
\left[\begin{array}{cc}S & 0 \\
0 & -S^\#\end{array}\right]; \nonumber \\
 H &=&  N; \nonumber \\
 K &=& \left[\begin{array}{cc}S & 0 \\
0 & S^\#\end{array}\right].
\end{eqnarray}

The QSDEs (\ref{qsde3}), (\ref{FGHKform})
define the general class of linear quantum systems considered. Such quantum systems can be used to model a large range of
devices and networks of devices arising in the area of quantum optics
including optical cavities, squeezers, optical parametric amplifiers,
cavity QED systems, 
beam splitters, and  phase shifters; e.g., see
\cite{DJ99,YK03a,YK03b,JNP1,MAB08,PET08A,NJD09,WM10,WM94,GZ00,BR04,GW09}.

\subsection{Coherent Quantum $H^\infty$ Control}
\label{sec:Hinf}
We formulate a coherent quantum control problem in
which a linear quantum system is controlled by a feedback controller
which is itself a linear quantum system. Such a controller is said to be physically realizable; e.g., see \cite{JNP1,PET10Ba,PET13Da}. The fact that the controller
is to be a quantum system means that any controller synthesis method
needs to produce controllers which are physically realizable.  The
 problem we consider is the quantum $H^\infty$ control problem in
which it is desired to design a coherent controller such that the
resulting closed loop quantum system is stable and attenuates
specified disturbances acting on the system; see
\cite{JNP1,MaP4}. In the standard quantum $H^\infty$ control
problem such as considered in \cite{JNP1,MaP4}, the quantum
noises are averaged out and only the external disturbance is
considered.  
 We now formulate the coherent quantum $H^\infty$ control
problem for a general class of quantum systems of the form
(\ref{qsde3}), (\ref{FGHKform}). 

We consider quantum \emph{plants} described by linear complex quantum stochastic
models of the
following form defined in an analogous way to the QSDEs
(\ref{qsde3}), (\ref{FGHKform}):

\begin{eqnarray}
\label{plant1}
\left[\begin{array}{l} d a(t)\\d a(t)^\#\end{array}\right]
 &=& F \left[\begin{array}{l}  a(t)\\
     a(t)^\#\end{array}\right]dt\nonumber \\
&&+\left[
{\begin{array}{*{20}c}
   {G_{0 } } & {G_{1 } } & {G_{2}}  \\
\end{array}} \right]\left[ {\begin{array}{*{20}c}
   {dv\left( t \right)}  \\  {dw \left( t \right)} \\ {du\left( t \right)}
\end{array}} \right];  \nonumber\\
dz \left( t \right) &=& H_{1 }\left[\begin{array}{l}  a(t)\\
     a(t)^\#\end{array}\right] dt 
+ K_{{12} } du \left( t \right); \nonumber\\
 dy \left( t \right) &=& H_{2 } 
\left[\begin{array}{l}  a(t)\\
     a(t)^\#\end{array}\right] dt \nonumber \\
&&+ \left[ {\begin{array}{*{20}c}
   {K _{{20} } } & {K _{{21} } } & {0 }  \\
\end{array}} \right] \left[ {\begin{array}{*{20}c}
  {dv \left( t \right)}  \\ {dw \left( t \right)}\\ {du \left( t \right)}
\end{array}} \right]\nonumber \\
\end{eqnarray}
where all of the matrices in these QSDEs have a form as in
(\ref{FGHKform}). 
Here, the input 
\[
dw(t)=\left[\begin{array}{l} \beta_w (t)dt+d\mathcal{A}(t)
\\ \beta_w^\# (t)dt+ d\mathcal{A}(t)^{\#} \end{array}\right]
\]
represents a disturbance signal where $\beta_{w}(t)$
is an adapted process; see \cite{JNP1,MaP3,PAR92}. The signal
$u(t)$ is a control input of the form
\[
du(t)=\left[\begin{array}{l} \beta_u (t)dt+d\mathcal{B}(t)
\\ \beta_u^\# (t)dt+ d\mathcal{B}(t)^{\#} \end{array}\right]
\] 
where  $\beta_{u}(t)$ is an adapted process. The
quantity 
\[
dv(t) = \left[\begin{array}{l} d\mathcal{C}(t)
\\ d\mathcal{C}(t)^{\#} \end{array}\right]
\] 
represents
any additional quantum noise in the plant. The quantities 
 $\left[\begin{array}{l} d\mathcal{A}(t)
\\ d\mathcal{A}(t)^{\#} \end{array}\right]$,
$\left[\begin{array}{l} d\mathcal{B}(t)
\\ d\mathcal{B}(t)^{\#} \end{array}\right]$ and $\left[\begin{array}{l} d\mathcal{C}(t)
\\ d\mathcal{C}(t)^{\#} \end{array}\right]$ are quantum noises.

In the coherent quantum $H^\infty$ control problem, we consider
controllers which are described by QSDEs of the form 
(\ref{qsde3}), (\ref{FGHKform}) as follows:
\begin{eqnarray}
\label{controller1}
\left[\begin{array}{l} da_c(t)\\da_c(t)^\#\end{array}\right]
 &=& F_c \left[\begin{array}{l} a_c(t)\\
     a_c(t)^\#\end{array}\right]dt\nonumber \\
&&+ \left[\begin{array}{*{20}c}
{\bar G_c} & \hat G _{c} & {G_{c}}
\end{array} \right] 
\left[\begin{array}{*{20}c}
   {d\bar w_{c} }  \\
   {d\hat w_{c} }  \\
   {dy}
\end{array}\right]
  \nonumber\\
\left[\begin{array}{c}
du(t) \\
d \hat u(t) \\
d \bar u(t)
\end{array}\right]
 &=& 
\left[\begin{array}{c}
H_c \\
\hat H_{c} \\
\bar H_{c}
\end{array}\right]
\left[\begin{array}{l} a_c(t)\\
     a_c(t)^\#\end{array}\right] dt  \nonumber \\
&&+\left[\begin{array}{ccc}
K_c & 0 & 0 \\
0 & \hat K_{c} & 0 \\
0 & 0 & \bar K_{c}
\end{array}\right]
\left[\begin{array}{*{20}c}
   {d\bar w_{c} }  \\
   {d\hat w_{c} }  \\
   {dy}
\end{array}\right]\nonumber \\
\end{eqnarray}
where all of the matrices in these QSDEs have a form as in
(\ref{FGHKform}). Here the quantities 
 \[
d\bar w_{c} = \left[\begin{array}{l} d\mathcal{A}_c(t)
\\ d\mathcal{A}_c(t)^{\#} \end{array}\right],~~
d\hat w_{c}=\left[\begin{array}{l} d\mathcal{B}_c(t)
\\ d\mathcal{B}_c(t)^{\#} \end{array}\right]
\]
  are controller quantum
noises. Also, the
outputs $du_0$ and $du_1$ are unused outputs of the controller which
have been included so that the controller can be  physically realizable. 
Corresponding to the plant (\ref{plant1}) and (\ref{controller1}), we
form the closed loop quantum system by identifying the output of the
plant $dy$ with the input to the controller $dy$, and identifying the
output of the controller $du$ with the input to the plant $du$. This
leads to the following closed-loop  QSDEs:
\begin{eqnarray} 
\label{closed1}
  d\eta  \left( t \right) &=& \left[ {\begin{array}{*{20}c}
    {F } & {G _{2 } H _{c } }  \\
    {G _{c } H _{2 } } & {F _{c } }  \\
 \end{array}} \right]\eta \left( t \right)dt \nonumber \\
&&+\left[ {\begin{array}{*{20}c}
    {G _{0 } } & {G _{2 }} & 0  \\
    {G _{c } K _{{20} } } & {\bar G_c} & {\hat G_c}\\
 \end{array}} \right]\left[ {\begin{array}{*{20}c}
    {dv \left( t \right)}  \\
    {d\bar w_{c} \left( t \right)}  \\
    {d\hat w_{c}\left( t \right)}  \\
 \end{array}} \right] \nonumber \\
&&+ \left[ {\begin{array}{*{20}c}
    {G _{1 } }  \\
    {G _{c } K _{{21} } }  \\
 \end{array}} \right]dw \left( t \right);\nonumber \\
   dz \left( t \right) &=& \left[ {\begin{array}{*{20}c}
    {H _{1 } } & {K _{{12} } H _{c } }  \\
 \end{array}} \right]\eta  \left( t \right)dt \nonumber \\
&&+ \left[ {\begin{array}{*{20}c}
    0 & {K _{{12} } }& 0  \\
 \end{array}} \right]\left[ {\begin{array}{*{20}c}
   {dv \left( t \right)}  \\
    {d\bar w_{c} \left( t \right)}  \\
    {d\hat w_{c}\left( t \right)}  \\
\end{array}} \right]
\end{eqnarray}
where
\[
\eta \left( t \right) = \left[\begin{array}{l} a(t)\\
 a(t)^\#\\ a_c(t)\\     a_c(t)^\#\end{array}\right].
\]

For a given quantum plant of the form (\ref{plant1}), the coherent quantum
$H^\infty$ control problem involves finding a physically realizable
quantum controller (\ref{controller1}) such that the resulting closed
loop system (\ref{closed1}) is such that the following conditions are
satisfied:
\begin{enumerate}[(i)]
\item
The matrix
\begin{equation}
\label{cl_hurwitz}
F_{cl} = \left[ {\begin{array}{*{20}c}
    {F } & {G _{2 } H _{c } }  \\
    {G _{c } H _{2 } } & {F _{c } }  \\
 \end{array}} \right]
\end{equation}
is Hurwitz;
\item
 The closed loop transfer function 
\[
\Gamma_{cl}(s) = H_{cl}\left(sI-F_{cl}\right)^{-1}G_{cl}
\]
satisfies
\begin{equation}
\label{cl_hinf}
\|\Gamma_{cl}(s)\|_\infty < 1
\end{equation}
where
\[
H_{cl} = \left[ {\begin{array}{*{20}c}
    {H _{1 } } & {K _{{12} } H _{c } }  \\
 \end{array}} \right], 
~~ G_{cl} = \left[ {\begin{array}{*{20}c}
    {G _{1 } }  \\
    {G _{c } K _{{21} } }  \\
 \end{array}} \right].
\]
\end{enumerate}

\begin{theorem}(See also \cite{JNP1,PAJ91}.)
\label{T3}
Suppose that the plant (\ref{plant1}) satisfies the following conditions:
\begin{enumerate}[(i)]
\item
$E_1 = K_{12}^\dagger K_{12} > 0$;
\item
$E_2 = K_{21} K_{21}^\dagger > 0$;
\item
The matrix $\left[\begin{array}{cc}F-i\omega I & G_2\\H_1 & K_{12}\end{array}\right]$ is full rank for all $\omega$;
\item
The matrix $\left[\begin{array}{cc}F-i\omega I & G_1\\H_2 & K_{21}\end{array}\right]$ is full rank for all $\omega$.
\end{enumerate}
Then the above coherent $H^\infty$ control problem has a solution if and only if the Riccati equations
\begin{eqnarray}
\label{r1} &&(F-G_2E_1^{-1}K_{12}^\dagger H_1)^\dagger X
+X(F-G_2E_1^{-1}K_{12}^\dagger H_1)+ \nonumber\\
&&\quad X(G_1G_1^\dagger  - G_2E_1^{-1}G_2')X+\nonumber\\
&&\qquad H_1^\dagger (I-K_{12}E_1^{-1}K_{12}^\dagger )H_1 = 0;
\end{eqnarray}

\begin{eqnarray}
\label{r2}&&(F-G_1K_{21}^\dagger E_2^{-1}H_2)Y+Y(F-G_1K_{21}^\dagger E_2^{-1}H_2)
+\nonumber\\
&&\quad Y(H_1^\dagger H_1-H_2^\dagger E_2^{-1}H_2)Y+\nonumber\\
&&\qquad G_1(I-K_{21}^\dagger E_2^{-1}K_{21})G_1^\dagger  = 0.
\end{eqnarray}
have positive-semidefinite stabilizing solutions $X$ and $Y$ such that the spectral radius of $XY$ is strictly less than one. In this case, the matrices $F_c$, $G_c$, and $H_c$ in the controller system (\ref{controller1}) can be constructed according to the formulas:
\begin{eqnarray}
\label{controller_matrices}
F_c &=& F+G_2H_c-G_cH_2+(G_1-G_cK_{21})G_1^\dagger X; \nonumber \\
G_c &=& (I-YX)^{-1}(YH_2^\dagger +G_1K_{21}^\dagger )E_2^{-1}; \nonumber \\
H_c&=& -E_1^{-1}(G_2^\dagger X+K_{12}^\dagger  H_1).
\end{eqnarray}
\end{theorem}

\subsection{Riccati Equations in the Physical Realizability of Linear Quantum Systems}
In the coherent quantum $H^\infty$ control problem considered above (see also, \cite{JNP1,MaP4,PET10B,PET13Da} as well in other coherent quantum control problems such as the coherent quantum LQG problem \cite{NJP1}), it is required that the controller be a physically realizable quantum system. One way to achieve this is to first design a classical linear controller and then to add additional quantum noises to the controller to make it physically realizable; e.g., see \cite{JNP1,PET13Da,VuP5}. In this section, we consider a result from \cite{VuP5} in which the Riccati equation plays a key role in this problem when it is desired to minimize the number of added quantum noises to make a system physically realizable. 

Here, we consider quantum linear systems described by QSDEs of the following form (see ~\cite{JNP1}):
\begin{eqnarray}
	dx(t) &=& A x(t) d t + B dw(t) 
		; \nonumber \\
	dy(t) &=& C x(t) d t + D d w(t). 
	\label{eqn:model}
\end{eqnarray}
Here, $x(t) = \begin{bmatrix}x_1(t) & \cdots & x_n(t)\end{bmatrix}^T$
is a column vector of $n$ self-adjoint system variables which are operators on the underlying 
Hilbert space. The components of vector $x$ are either position operators $q_i$ or momentum operators $p_i$, which are related to the annihilation and creation operators considered above as follows:
\[
q_i = a_i + a_i^*; \quad p_i = -\imath a_i + \imath a_i^*.
\]
Similarly, 
$d w(t)$ is a column vector of $n_w$ self-adjoint, non-commutative operators 
representing the input to the system and  
$d y(t)$ is a column vector of $n_y$ self-adjoint, non-commutative operators 
representing the output of the system.
 The components of vector $w$ are either input field position operators $\mathcal{Q}_i$ or input field momentum operators $\mathcal{P}_i$, which are related to the input field annihilation and creation operators considered above as follows:
\[
\mathcal{Q}_i = \mathcal{A}_i + \mathcal{A}_i^*; \quad \mathcal{P}_i = -\imath \mathcal{A}_i + \imath \mathcal{A}_i^*.
\]
Also, the components of the vector $y$ are either output field position operators $\mathcal{Q}_i^{out}$ or output field momentum operators $\mathcal{P}_i^{out}$, which are related to the output field annihilation and creation operators considered above as follows:
\[
\mathcal{Q}^{out}_i = \mathcal{A}^{out}_i + \mathcal{A}^{out *}_i; \quad \mathcal{P}^{out}_i = -\imath \mathcal{A}^{out}_i + \imath \mathcal{A}^{out *}_i.
\]

We consider the problem of implementing an arbitrary, strictly proper, 
LTI system as a quantum system 
(for example when implementing a coherent controller) by introducing 
vacuum noise sources. The resulting quantum 
systems are described by the following QSDEs 
which are a special case of (\ref{eqn:model}):
\begin{eqnarray}
	d x(t) &=& A x(t) d t + B_u d u(t) \nonumber \\
	&& {} + B_{v_1} d v_1(t) +  B_{v_2} d v_2(t)  
		; \nonumber \\
	d y(t) &=& C x(t) d t + d v_1(t).
	\label{eqn:model2}
\end{eqnarray}

Here, $ u(t)$ (a column vector with $n_u$ components) 
represents the inputs to the system.
Also, $ v_1(t)$ and $ v_2(t)$ (column vectors with $n_{v_1}$ and $n_{v_2}$ components respectively) 
are quantum Wiener processes 
corresponding to the introduced vacuum noise inputs. For convenience, the vacuum noises are 
partitioned into two vectors $ v_1(t)$ and $ v_2(t)$ such that $n_{v_1} = n_u$. 
Then, $n_v = n_{v_1} + n_{v_2}$ is the total number of introduced vacuum noise inputs. 
Subsequently, we will refer to $ v_1$ as the \emph{direct feedthrough quantum noises} and to 
$ v_2$ as the \emph{additional quantum noises}.

\begin{definition} \label{def:pr}
	The system described by (\ref{eqn:model}) is \emph{physically realizable} if 
	there exists 
	a quadratic Hamiltonian operator $\mathcal{H} = \frac{1}{2} x(0)^TRx(0)$,
	where $R$ is a real, symmetric, $n \times n$ matrix, and  
	a coupling operator vector $\mathcal{L} = \Lambda x(0)$, 
	where $\Lambda$ is a complex-valued $\frac{1}{2} n_w \times n$ 
	coupling matrix  
	such that the matrices $A$, $B$, $C$ and $D$ are given by:
	\begin{subequations}
	\begin{align}
		A &= 2 \Theta \left(R + \mathfrak{Im}\left(
			\Lambda^{\dagger}\Lambda \right) \right);
			\label{eqn:a} 
			\\
		B &= 2i \Theta \begin{bmatrix}
			-\Lambda^{\dagger} & \Lambda^T \end{bmatrix}\Gamma; 
			\label{eqn:b} 
			\\
		C &= P^T \begin{bmatrix}
			\Sigma_{n_y} & 0 \\ 0 & \Sigma_{n_y} \end{bmatrix}
			\begin{bmatrix} \Lambda + \Lambda^\# \\
				-i\Lambda + i\Lambda^\# \end{bmatrix};
			\label{eqn:c} 
			\\
		D &= \begin{bmatrix} I_{n_y \times n_y} &
			0_{n_y \times \left(n_w - n_y\right)} 
			\end{bmatrix}.
			\label{eqn:d} 
	\end{align}
	\end{subequations}
	Here: 
	\begin{equation}
	\Theta = \begin{bmatrix} 
	J & 0 & \cdots & 0 \\
	0 & J & \cdots & 0 \\
	\vdots  & \vdots & \ddots & \vdots \\
	0 & 0 & \cdots & J
	\end{bmatrix};
	\qquad 
	J = \begin{bmatrix}0 & 1 \\ -1 & 0
	\end{bmatrix}.   
		\label{eqn:ctheta}
	\end{equation}	
$\Gamma_{n_w \times n_w} = P \diag (M)$; 
		$M = \frac{1}{2}\left[ \begin{smallmatrix}1 & i \\ 1 & -i 
			\end{smallmatrix} \right]$;  
		$\Sigma_{n_y} = \begin{bmatrix}
			I_{\frac{1}{2}n_y \times \frac{1}{2}n_y} &
			0_{\frac{1}{2}n_y \times \frac{1}{2}\left(
				n_w - n_y \right) } \end{bmatrix}$;  
	$P$ is the appropriately dimensioned square permutation 
	matrix such that 
	$ P \begin{bmatrix}a_1 & a_2 & \cdots & a_{2m} \end{bmatrix} $ \linebreak
$	=\begin{bmatrix}a_1 & a_3 & \cdots & a_{2m-1} 
	a_2 & a_4 & \cdots & a_{2m} \end{bmatrix}$
	and 
	$\diag (M)$ is an appropriately dimensioned square block diagonal 
	matrix with each diagonal block equal to the matrix $M$. (Note that the  
	dimensions of $P$ and $\diag (M)$ can always be determined from the
	context in which they appear.) $\mathfrak{Im}\left(.\right)$ 
	denotes the imaginary part of a matrix and ${}^\dagger$ denotes the 
	complex conjugate transpose of a matrix.
\end{definition}

The following theorem shows how a Riccati equation with a skew-symmetric solution can be used to determine if a given transfer function can be physically realized with only direct feedthrough noises and no additional quantum noises. 

\begin{theorem}
	\label{thm:tf}
	Consider a system with strictly proper transfer function matrix: 
	$$G(s) = \tilde{C}(sI - \tilde{A})^{-1}\tilde{B}_u.$$ 
	Suppose the algebraic Riccati equation (ARE) 
	\begin{equation}
		X \tilde{B}_u \Theta_{n_u} \tilde{B}_u^T X 
		- \tilde{A}^T X - X \tilde{A}	
		- \tilde{C}^T \Theta_{n_y} \tilde{C} = 0 
		\label{eqn:ric1}
	\end{equation}
	has a non-singular, real, skew-symmetric solution $X$. 
	Here, the matrices $\Theta_{n_u}$ and $\Theta_{n_y}$ are 
	defined as in (\ref{eqn:ctheta}). 
	Then there exists matrices $\left\{ A, B_u, C \right\}$ 
	such that 
	$$G(s) = C(sI - A)^{-1}B_u$$
	and the corresponding system (\ref{eqn:model2}) is physically realizable 
	with only the direct feedthrough quantum noises $ v_1$ and no 
	additional quantum noises $ v_2$. 
\end{theorem}

%\bibliography{C:/Users/Ian/irp/Bibliog/irpnew}
% \bibliography{/home/irp/Bibliog/irpnew}
% \bibliographystyle{IEEEtran}

\end{document}